\documentclass[conference,a4paper,romanappendices,10pt]{IEEEtran}	

\IEEEoverridecommandlockouts

\usepackage{algorithm,algorithmic,amsmath,amssymb,amsthm,array,bibentry,cite,color,comment,enumerate}
\usepackage{enumitem,eurosym,float,graphicx,lettrine,mathrsfs,multicol,multirow,nomencl,pict2e,psfrag}
\usepackage{ragged2e,setspace,subfigure,supertabular,tabularx,url}
\usepackage[USenglish]{babel}

\setlist[itemize]{leftmargin=12.5mm}

{		}
{ 		}
{ 		}
{ 		}
{ 		}
{		}
{		}
{ 		}

\newcommand{\h}{\mathbf{h}}

\newcommand{\n}{\mathbf{n}}

\renewcommand{\v}{\mathbf{v}}
\newcommand{\w}{\mathbf{w}}

\newcommand{\y}{\mathbf{y}}

\renewcommand{\H}{\mathbf{H}}

\newcommand{\setB}{\mathcal{B}}
\newcommand{\setC}{\mathcal{C}}
\newcommand{\setD}{\mathcal{D}}

\newcommand{\setN}{\mathcal{N}}

\newcommand{\setU}{\mathcal{U}}

\newcommand{\Real}{\mbox{$\mathbb{R}$}}
\newcommand{\Compl}{\mbox{$\mathbb{C}$}}

\newcommand{\argmax}{\operatornamewithlimits{argmax}}

\newcommand{\herm}{\mathrm{H}}

\newcommand{\DL}{\textnormal{\tiny{DL}}}
\newcommand{\UL}{\textnormal{\tiny{UL}}}

\begin{document}

\title{Title}

\title{Performance Evaluation of User Scheduling for Full-Duplex Small Cells in Ultra-Dense Networks}

\author{
\IEEEauthorblockN{Italo Atzeni, Marios Kountouris, and George C. Alexandropoulos}
\IEEEauthorblockA{Mathematical and Algorithmic Sciences Lab, France Research Center, Huawei Technologies Co$.$ Ltd$.$}
emails: \{italo.atzeni, marios.kountouris, george.alexandropoulos\}@huawei.com}
\maketitle

\begin{abstract}
Full-duplex (FD) communication is an emerging technology that can potentially double the throughput of cellular networks. Preliminary studies in single-cell or small FD network deployments have revealed promising rate gains using self-interference cancellation (SIC) techniques and receive processing. Nevertheless, the system-level performance gains of FD small cells in ultra-dense networks (UDNs) have not been fully investigated yet. In this paper, we evaluate the performance of resource allocation in ultra-dense FD small-cell networks using spatial stochastic models for the network layout and 3GPP channel models. More specifically, we consider various UDN scenarios and assess the performance of different low-complexity user-scheduling schemes and power allocation between uplink and downlink. We also provide useful insights into the effect of the SIC capability on the network throughput.
\end{abstract}

\vspace{1mm}


\section{Introduction} \label{sec:Intro}

\noindent Full-duplex (FD) communication has the potential to cope with the ever growing demand for high data rates. FD systems can -- at least theoretically -- double the system throughput by enabling simultaneous transmission and reception in the same time/frequency resource \cite{Sab14}. Significant progress has been made in addressing the two major challenges hindering the implementation of FD systems, namely the self-interference (SI) at the base station (BS), due to signal leakage, and the inter-user interference in the downlink (DL), due to concurrent uplink (UL) transmission at the same time/frequency. Prior work has demonstrated important throughput gains in single-cell FD systems or in network scenarios with small number of BSs, either theoretically or using experimental/simulation-based results. A major current research focus is on the impact of the residual SI on the overall performance \cite{Atz16}. 

Although resource allocation in half-duplex (HD) systems has been extensively studied, the new characteristics and challenges of resource management in FD need further analysis \cite{Son15}. In fact, without carefully allocating resources and selecting users, FD transmission may cause excessive interference in both UL and DL, which may greatly limit the potential FD gains \cite{Goy15}. In this regard, \cite{Wan15} presents a distributed power allocation scheme for multi-cell FD scenarios that aims at maximizing the network throughput. A dynamic power control scheme in FD bidirectional networks was provided in \cite{Che13}. In \cite{Sul15,Nam15}, the joint resource allocation problem of user pairing, subcarrier allocation, and power control for FD orthogonal frequency division multiple access (OFDMA) networks was considered. Two low-complexity user pairing algorithms that maximize the cell throughput or minimize the outage probability were presented in \cite{Cho14}. The performance of FD small cells in ultra-dense networks (UDNs) has been studied in \cite{Atz15a} using tools from stochastic geometry; however, analyzing user scheduling in such framework remains a major challenge.

\begin{figure}[t!]
\centering
\includegraphics[scale=0.97]{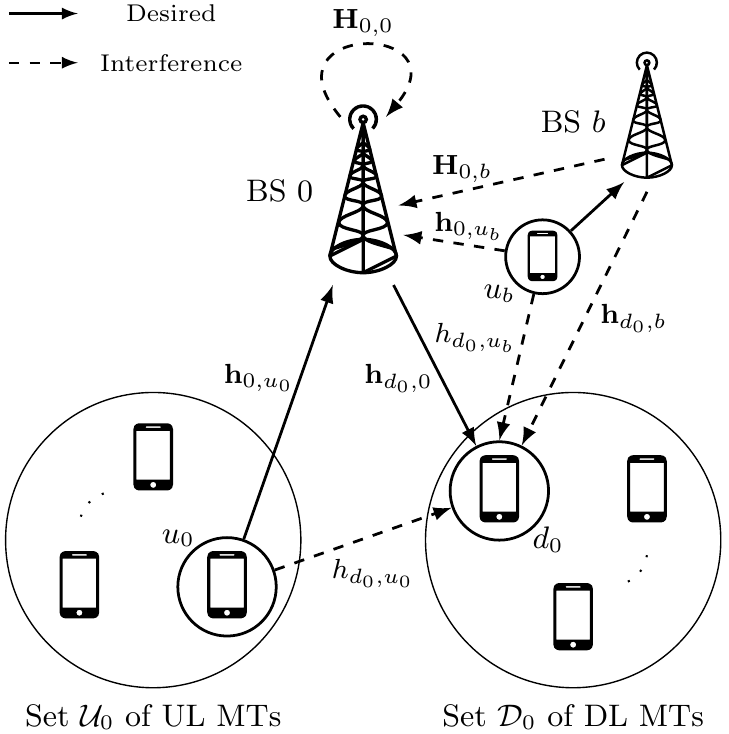} 
\caption{Reference FD small cell with desired and interfering channels.} \label{fig:scheme} \vspace{-3mm}
\end{figure}

\begin{figure*}
\begin{align} 
\label{eq:y_0} \y_{0} & \triangleq \sqrt{p_{\UL} \ell(r_{0,u_{0}})} \h_{0,u_{0}} s_{u_{0}} + \sqrt{p_{\DL}} \H_{0,0} \w_{0} s_{0} + \sum_{b \in \setB} \Big( \sqrt{p_{\DL} \ell(r_{0,b})} \H_{0,b} \w_{b} s_{b} + \sqrt{p_{\UL} \ell(r_{0,u_{b}})} \h_{0,u_{b}} s_{u_{b}} \Big) + \n_{0} \\
\label{eq:y_d} y_{d_{0}} & \triangleq \sqrt{p_{\DL} \ell(r_{d_{0},0})} \h_{d_{0},0}^{\herm} \w_{0} s_{0} \! + \! \sqrt{p_{\UL} \ell(r_{d_{0},u_{0}})} h_{d_{0},u_{0}} s_{u_{0}} \! + \sum_{b \in \setB} \Big( \sqrt{p_{\DL} \ell(r_{d_{0},b})} \h_{d_{0},b}^{\herm} \w_{b} s_{b} \! + \! \sqrt{p_{\UL} \ell(r_{d_{0},u_{b}})} h_{d_{0},u_{b}} s_{u_{b}} \Big) \! + n_{d_{0}}
\end{align}
\hrulefill \vspace{-2mm}
\end{figure*}

In this work, we consider the problem of user scheduling and power allocation for ultra-dense FD small-cell networks. UDNs are another enabling technology for achieving higher data rates and enhanced coverage by exploiting spatial reuse, while retaining at the same time seamless connectivity and mobility of cellular networks. Inspired by the attractive features and potential advantages of UDNs, their development and deployment has gained momentum in both the wireless industry and the research community during the last years. It has also attracted the attention of standardization bodies such as 3GPP LTE-Advanced. Using spatial stochastic models to represent the BS locations, namely hardcore point processes (HCPP), and incorporating 3GPP channel models and configurations, we evaluate the performance of three low-complexity user-scheduling algorithms with and without optimal power allocation (OPA) \cite{Ale16}. Our framework allows us to quantify the impact of interference and to identify the dominant interference terms in different scenarios. Our simulation results show the significant sum-rate gains of FD small cells in large network deployments and in different operating regimes. We also provide insights into the impact of the SI on the resource allocation and the network throughput performance.

\section{System Model} \label{sec:SM}

We consider the scenario illustrated in Figure~\ref{fig:scheme}, in which a reference small-cell FD BS $0$ communicates with a set $\setU_0$ of HD mobile terminals (MTs) in the UL and with a set $\setD_0$ of HD MTs in the DL. BS $0$ is part of a dense FD small-cell network consisting of additional BSs, which are grouped in the set $\setB$. In our network setup, all FD BSs are equipped with $N_{T}$ transmit and $N_{R}$ receive antennas, whereas each MT has a single antenna. During a given time-frequency resource unit, BS $0$ schedules concurrently one MT from $\setU_0$ and one MT from $\setD_0$ so as to maximize the sum-rate performance; the selected UL and DL MTs are denoted by $u_0\in\setU_0$ and $d_0\in\setD_0$, respectively. The same procedure takes place for every BS $b\in\setB$. For example, in Figure~\ref{fig:scheme}, we sketch the selected UL MT from BS $b\in\setB$, denoted as $u_b\in\setU_b$, with $\setU_{b}$ representing the set of UL MTs associated with BS $b$. Both UL MT $u_b$ and BS $b$ create interference at the reference small cell. The transmit powers of each BS and of each UL MT in the network are denoted as $p_{\DL} \in [0, P_{\DL}]$ and $p_{\UL} \in [0, P_{\UL}]$, respectively, where $P_{\DL}$ and $P_{\UL}$ represent the maximum power for DL and UL communication, respectively. 

The wireless propagation channels are assumed to be subject to a combination of large-scale propagation effects, mainly including pathloss attenuation, shadowing, and small-scale fading. We consider the case of distance-dependent pathloss between a receiving node $j$ and a transmitting node $k$, given by the function $\ell(r_{j,k})$, where $r_{j,k} \triangleq \|j - k\|$ is the Euclidean distance between these two nodes. Focusing on the reference FD small cell in Figure~\ref{fig:scheme}, $\H_{0,b} \in \Compl^{N_{R} \times N_{T}}$ represents the channel between BSs $0$ and $b$, and $\h_{0,k} \in \Compl^{N_{R}}$ denotes the channel between BS $0$ and UL MT $k$ with $k \in \setU_{0} \cup (\cup_{b \in \setB} \setU_{b})$. In addition, $\H_{0,0} \in \Compl^{N_{R} \times N_{T}}$ models the residual SI channel seen at the receive antennas of BS $0$ due to its own DL transmission. The channel between DL MT $d \in \setD_{0}$ and each BS $\ell \in \{0\} \cup \setB$ is represented by $\h_{d,\ell} \in \Compl^{N_{T}}$,  whereas $h_{d,k} \in \Compl$ denotes the inter-MT interference channel from the $k$-th UL MT to the $d$-th DL MT.

We assume that each FD BS $\ell$ has perfect channel state information (CSI) for its selected UL and DL MTs and, assuming that CSI about the inter-cell interference is not available, it performs matched filtering at both the desired received and transmitted signals. More specifically, each BS $\ell$ utilizes the maximum ratio combining (MRC) vector $\v_{\ell} \in \Compl^{N_{R}}$ and the maximum ratio transmission (MRT) vector $\w_{\ell} \in \Compl^{N_{T}}$, with $\| \v_{\ell} \|^{2} = \| \w_{\ell} \|^{2} = 1$. For the reference FD small cell in Figure~\ref{fig:scheme}, the baseband received signal at BS $0$ and at scheduled DL MT $d_{0}$ can be mathematically expressed as in \eqref{eq:y_0} and \eqref{eq:y_d}, respectively, shown at the top of the page. Within these expressions, $s_{k}$ with $k=\{0,u_0,b,u_b\}$ represents the unit-power data symbol transmitted by node $k$, $\n_{0} \in \Compl^{N_{R}}$ denotes the additive noise at BS $0$ with elements distributed independently as $\setC \setN (0, \sigma_{0}^{2})$, and $n_{d_{0}} \in \Compl$ is the additive noise at the scheduled DL MT $d_{0}$ distributed as $\setC \setN (0, \sigma_{d_{0}}^{2})$. From \eqref{eq:y_0} and \eqref{eq:y_d}, the received signal-to-interference-plus noise ratio (SINR) at BS $0$ after applying the combining vector $\v_{0}$ and the SINR at the scheduled DL MT $d_{0}$ are given by
\begin{align}
\label{eq:SINR_0} \mathsf{SINR}_{0} & \triangleq \frac{S_{0,u_{0}}}{I_{0,0} + I_{0,\setB} + I_{0,\setU_{B}} + \sigma_{0}^{2}} \\
\label{eq:SINR_d} \mathsf{SINR}_{d_{0}} & \triangleq \frac{S_{d_{0},0}}{I_{d_{0},u_{0}} + I_{d_{0},\setB} + I_{d_{0},\setU_{B}} + \sigma_{d_{0}}^{2}}
\end{align} \vspace{2mm}
respectively, where in \eqref{eq:SINR_0} we have used the notation
\begin{align}
\label{eq:S_0_u} S_{0,u_{0}} & \triangleq p_{\UL} \ell(r_{0,u_{0}}) |\v_{0}^{\herm} \h_{0,u_{0}}|^{2} \\
\label{eq:I_0_0} I_{0,0} & \triangleq p_{\DL} |\v_{0}^{\herm} \H_{0,0} \w_{0}|^{2} = \frac{p_{\DL}}{\Omega} \\
\label{eq:I_0_B} I_{0,\setB} & \triangleq p_{\DL} \sum_{b \in \setB} \ell(r_{0,b}) |\v_{0}^{\herm} \H_{0,b} \w_{b}|^{2} \\
\label{eq:I_0_U} I_{0,\setU_{B}} & \triangleq p_{\UL} \sum_{b \in \setB} \ell(r_{0,u_{b}}) |\v_{0}^{\herm} \h_{0,u_{b}}|^{2}
\end{align}
and in \eqref{eq:SINR_d} we have used the notation
\begin{align}
\label{eq:S_d_0} S_{d_{0},0} & \triangleq p_{\DL} \ell(r_{d_{0},0}) |\h_{d_{0},0}^{\herm} \w_{0}|^{2} \\
\label{eq:I_d_u} I_{d_{0},u_{0}} & \triangleq p_{\UL} \ell(r_{d_{0},u_{0}}) |h_{d_{0},u_{0}}|^{2} \\
\label{eq:I_d_B} I_{d_{0},\setB} & \triangleq p_{\DL} \sum_{b \in \setB} \ell(r_{d_{0},b}) |\h_{d_{0},b}^{\herm} \w_{b}|^{2} \\
\label{eq:I_d_U} I_{d_{0},\setU_{B}} & \triangleq p_{\UL} \sum_{b \in \setB} \ell(r_{d_{0},u_{b}}) |h_{d_{0},u_{b}}|^{2}
\end{align}
for the powers of the desired and interfering received signals. It is noted that \eqref{eq:I_0_0} and \eqref{eq:I_d_u} are the interference terms inside the reference small cell (namely, the SI and the inter-MT interference), whereas \eqref{eq:I_0_B}, \eqref{eq:I_0_U}, \eqref{eq:I_d_B}, and \eqref{eq:I_d_U} represent the inter-cell interference terms. Moreover, \eqref{eq:I_0_0}, \eqref{eq:I_0_B}, \eqref{eq:I_d_u}, and \eqref{eq:I_d_U} are the interference terms due to the FD mode of the BSs; if all BSs operated in HD mode, all these terms would vanish. The SI channel $\H_{0,0}$ can be modeled with Ricean distributed elements \cite{Dua12}: in this regard, a tight Gamma approximation for the distribution of the residual SI power $|\v_{0}^{\herm} \H_{0,0} \w_{0}|^{2}$ has been recently derived in \cite{Atz15a}. However, in this paper, we assume that $|\v_{0}^{\herm} \H_{0,0} \w_{0}|^{2}$ is constant and equal to $\Omega^{-1}$; this parameter is assumed to indicate the SIC capability at BS $0$ and, as such, it divides its DL transmit power as described in~\eqref{eq:I_0_0}.

\begin{figure*}[t]
\includegraphics[scale=0.98]{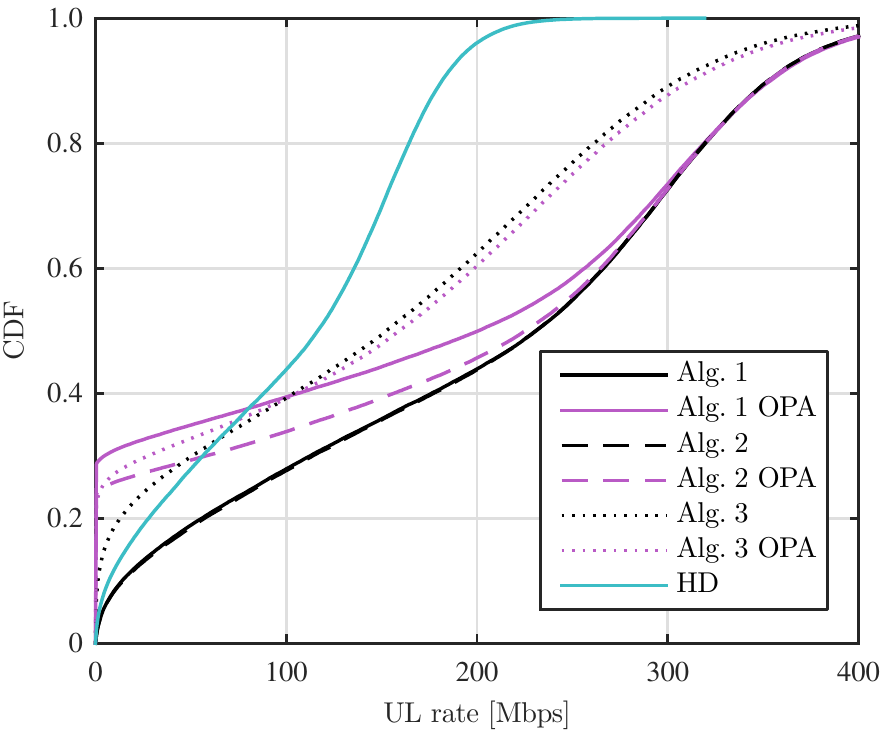} \hspace{3mm}
\includegraphics[scale=0.98]{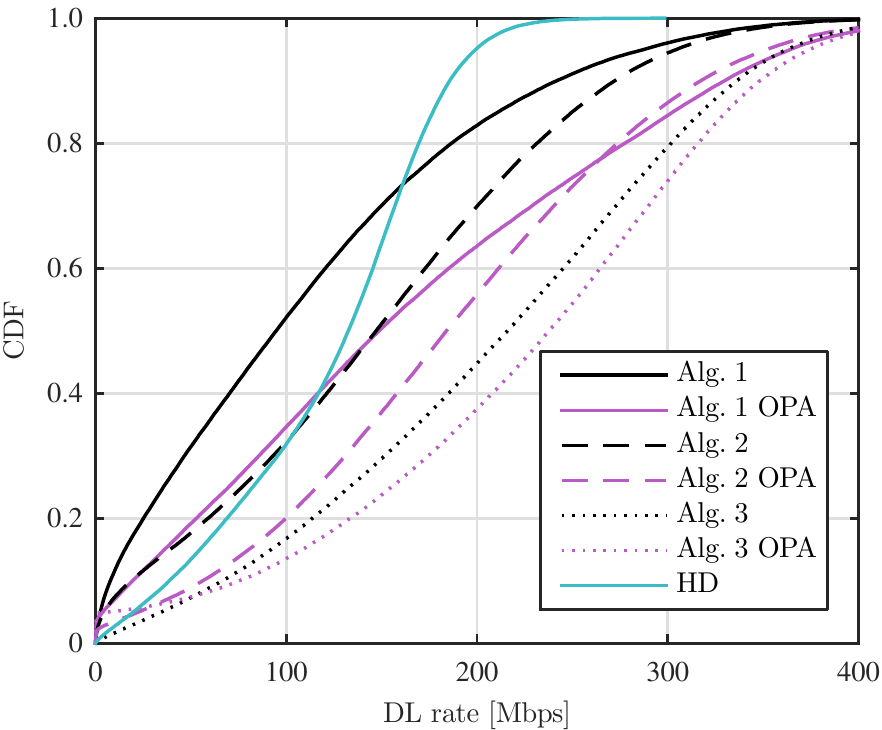}
\caption{Empirical CDFs of the UL and DL rate obtained with the different user-scheduling algorithms (with and without OPA), for $N_{T}=N_{R}=1$ and $\lambda=2.5 \times 10^{-5}$~BSs/m$^{2}$.} \label{fig:cdf_UL_DL} \vspace{-1mm}
\end{figure*}

Capitalizing on \eqref{eq:SINR_0} and \eqref{eq:SINR_d} for the reference small cell and assuming Gaussian signaling and ideal link adaptation, the instantaneous rates of the UL and DL transmissions, measured in [bps], are given by
\begin{align}
\label{eq:R_0} \mathsf{R}_{0} & \triangleq B \log_{2} (1 + \mathsf{SINR}_{0}) \\
\label{eq:R_d} \mathsf{R}_{d_{0}} & \triangleq B \log_{2} (1 + \mathsf{SINR}_{d_{0}})
\end{align}
respectively, where $B$ denotes the FD bandwidth. Finally, the sum rate of the FD system is the cumulative rate of the UL and DL transmissions, and is denoted as
\begin{align} \label{eq:R}
\mathsf{R} \triangleq \mathsf{R}_{0} + \mathsf{R}_{d_{0}}.
\end{align}

\section{User Scheduling and Power Allocation} \label{sec:User_Scheduling_Schemes}

In this section, we first consider the case where arbitrary transmit powers are utilized for the concurrent UL and DL transmissions and present three low-complexity user-scheduling algorithms for FD cellular networks \cite{Ale16}. Then, we provide the OPA strategy for maximizing the sum-rate performance.

\subsection{User-Scheduling Algorithms} \label{sec:Algos}

Let us assume fixed transmit powers $p_{\UL}$ and $p_{\DL}$; furthermore, we assume that all channel gains within the cell can be accurately estimated with an appropriately designed control plane, whereas the inter-cell interference terms are not known. In the following, we briefly present three user-scheduling algorithms for FD cellular networks, which are characterized by very low complexity and do not require any information exchange among BSs; we refer to \cite{Ale16} for further details on these algorithms.

\begin{itemize}
\item[Alg. 1)] UL MT $u_{0}$ and DL MT $d_{0}$ are selected as the ones with the highest channel gains in $\setU_{0}$ and in $\setD_{0}$, respectively:
\end{itemize}
\begin{align}
\label{eq:A1} \tag{$\mathrm{A1.1}$} u_{0} & = \argmax_{u \in \setU_{0}} S_{0,u} \\
\label{eq:A2} \tag{$\mathrm{A1.2}$} d_{0} & = \argmax_{d \in \setD_{0}} S_{d,0}.
\end{align}

\begin{itemize}
\item[Alg. 2)] UL MT $u_{0}$ is selected first as in \eqref{eq:A1}, and then DL MT $d_{0}$ is selected as the one experiencing the maximum SINR (neglecting the interference from the other cells):
\end{itemize}
\begin{align}
\label{eq:B2} \tag{$\mathrm{A2.1}$} d_{0} & = \argmax_{d \in \setD_{0}} \frac{S_{d,0}}{I_{d,u_{0}} + \sigma_d^{2}}.
\end{align}

\begin{itemize}
\item[Alg. 3)] DL MT $d_{0}$ is selected first as in \eqref{eq:A2}, and then UL MT $u_{0}$ is selected as the one yielding the maximum signal-to-leakage-plus-noise ratio (SLNR), i$.$e$.$, as the MT that simultaneously presents high UL channel gain and creates low interference to the scheduled DL MT $d_{0}$:
\end{itemize}
\begin{align}
\label{eq:C2} \tag{$\mathrm{A3.2}$} u_{0} & = \argmax_{u \in \setU_{0}} \frac{S_{0,u}}{I_{d_{0},u} + \sigma_{0}^{2}}.
\end{align}

Note that in Alg.~1, the steps of selecting UL MT $u_{0}$ and DL MT $d_{0}$ do not affect each other and, therefore, they may be performed independently. On the other hand, in Alg.~2 and Alg.~3, the CSI of the UL (respectively DL) MT channel gain is not required in the DL (respectively UL) scheduling step. In fact, once UL MT $u_0$ (respectively DL MT $d_0$) is selected, it suffices to know only $h_{d,u_0}$, $\forall$~$d \in \setD_{0}$ (respectively $h_{d_0,u}$, $\forall$~$u \in \setU_{0}$) for Alg.~2 (respectively for Alg.~3), instead of $h_{d,u}$, $\forall$~$d \in \setD_{0}$ and $\forall$~$u \in \setU_{0}$.

\subsection{Optimal Power Allocation (OPA)} \label{sec:PC}

The power allocation is assumed to take place after selecting $u_{0}$ and $d_{0}$ using one of the user-scheduling algorithms described in Section~\ref{sec:Algos}. It can be shown that the transmit powers $p_{\UL}^{\star}$ and $p_{\DL}^{\star}$ that maximize the sum-rate performance of the considered FD network are given by
\begin{align} \label{eq:max_sum-R}
(p_{\UL}^{\star}, p_{\DL}^{\star}) \triangleq \argmax_{\substack{p_{\UL} \in [0, P_{\UL}] \\ p_{\DL} \in [0, P_{\DL}]}} \mathsf{R}(p_{\UL}, p_{\DL}) = \argmax_{\substack{p_{\UL} \in \{0, P_{\UL}\} \\ p_{\DL} \in \{0, P_{\DL}\}}} \mathsf{R}(p_{\UL}, p_{\DL})
\end{align}
where we have highlighted the dependence of $\mathsf{R}$ on the transmit powers of the DL and UL communication. The second equality in \eqref{eq:max_sum-R} can be intuitively obtained by checking the convexity/concavity of the sum rate $\mathsf{R}$ with respect to transmit powers $p_{\UL}$ and $p_{\DL}$;  we address the interested reader to \cite{Ale16} for the detailed proof. Interestingly, the OPA strategy in \eqref{eq:max_sum-R} has binary feature and can be exploited in the scheduling procedure to maximize the network performance by switching between the optimal operation modes, i$.$e$.$, HD or FD mode. For the cases where the OPA strategy corresponds to HD mode, either in the UL or DL direction, one can repeat the scheduling of the UL or DL MT so as to maximize the HD rate: this corresponds to selecting the UL or DL MT as in \eqref{eq:A1} or \eqref{eq:A2}, respectively. This simple binary OPA can be incorporated in the scheduling procedure with the following additional step: 
\setlist[itemize]{leftmargin=10mm}
\begin{itemize}
\item[OPA)] After selecting UL MT $u_{0}$ and DL MT $d_{0}$ with any of the above user-scheduling algorithms, the FD/HD mode that maximizes the sum rate is determined as:
\begin{align} \label{eq:D2}
(p_{\UL}^{\star}, p_{\DL}^{\star}) = \argmax_{\substack{p_{\UL} \in \{0, P_{\UL}\} \\ p_{\DL} \in \{0, P_{\DL}\}}} \mathsf{R}(p_{\UL},p_{\DL}).
\end{align}
If the OPA yields HD mode in the UL (respectively DL), then UL MT $u_{0}$ (respectively DL MT $d_{0}$) is rescheduled so as to maximize the UL (respectively DL) rate.
\end{itemize}

\begin{figure}[t!]
\centering
\includegraphics[scale=0.98]{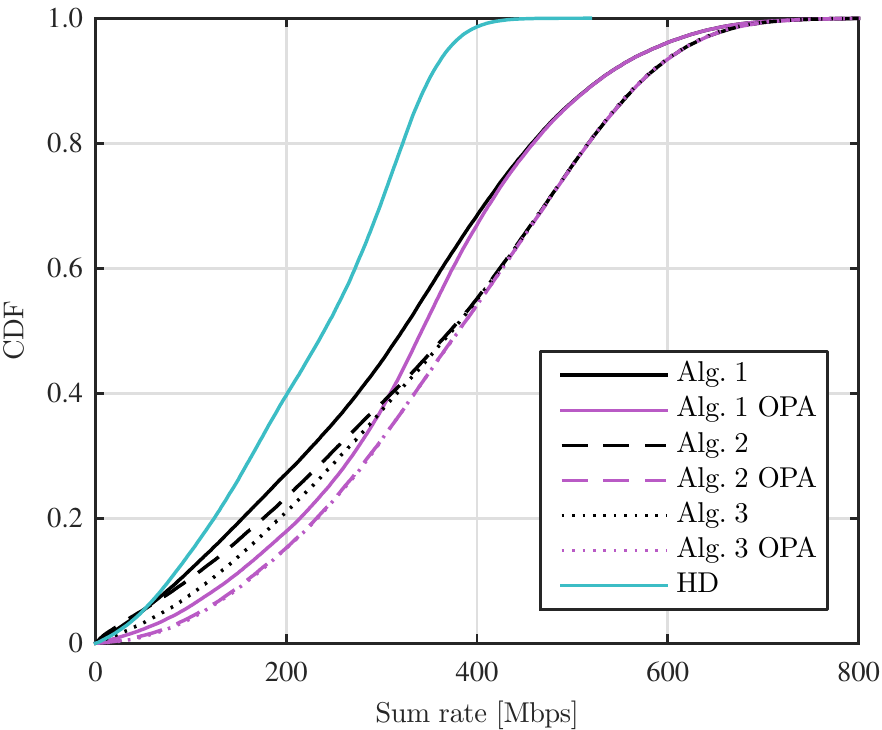}
\caption{Empirical CDF of the sum rate obtained with the different user-scheduling algorithms (with and without OPA), for $N_{T}=N_{R}=1$ and $\lambda=2.5 \times 10^{-5}$~BSs/m$^{2}$.} \label{fig:cdf_sum} \vspace{-1mm}
\end{figure}

\begin{figure}[t!]
\includegraphics[scale=0.98]{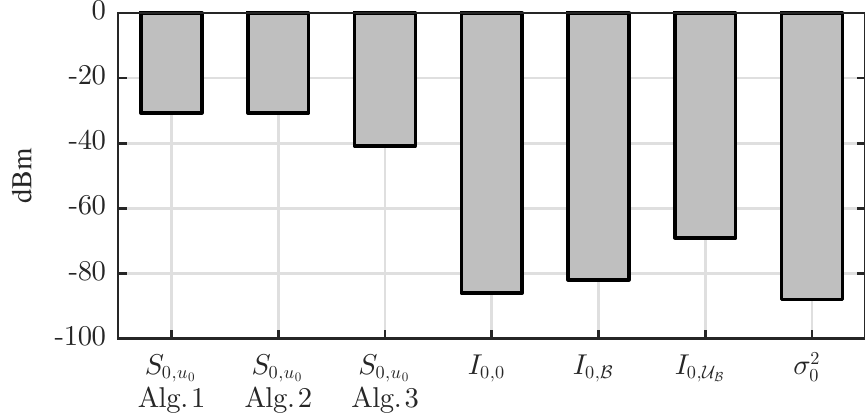} \\ \vspace{-2mm} \\
\includegraphics[scale=0.98]{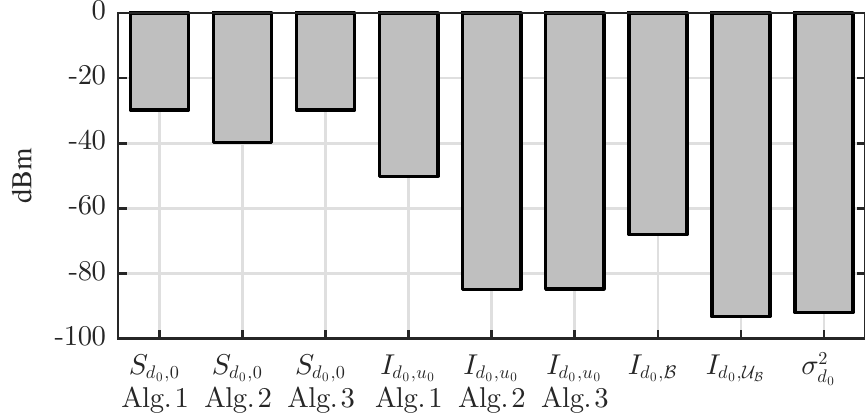}
\caption{Average contributions of the UL and DL SINRs obtained with the different user-scheduling algorithms, for $N_{T}=N_{R}=1$ and $\lambda=2.5 \times 10^{-5}$~BSs/m$^{2}$.} \label{fig:hist_UL_DL} \vspace{-1mm}
\end{figure}

\section{Numerical Results and Discussion} \label{sec:Num}

We assume that BS $0$ is located at the origin of the Euclidean plane, and that $U_{0} \triangleq |\setU_{0}|$ UL MTs and $D_{0} \triangleq |\setD_{0}|$ DL MTs are uniformly distributed within a distance $r_{0}$ around BS $0$. The distribution of the interfering FD BSs in $\setB$ and of their scheduled UL MTs in $\setU_{\setB}$ follows the marked Poisson HCPP $\Phi \triangleq \{(b, u_{b})\} \subset \Real^{2} \times \Real^{2}$. Here, the points $\{b\}$ have spatial density $\lambda$, measured in [BSs/m$^{2}$], and are forbidden to fall closer than a distance $2r_{0}$ from each other, thus preventing any overlap between MTs associated to different BSs;\footnote{HCPPs model real-world deployment of BSs more realistically than Poisson point processes; in addition, the pathloss model assumed by 3GPP TR 36.828 implicitly sets a minimum distance between BSs, preventing the interference coming from very close points to become unbounded.} on the other hand, the marks $\{u_{b}\}$ are such that $u_{b} = b + \sqrt{r_{b}} (\cos \varphi_{b}, \sin \varphi_{b})$, where $\{r_{b}\}$ and $\{\varphi_{b}\}$ are independent and uniformly distributed (i.i.d.) in $[0,1]$ and in $[0,2 \pi]$, respectively. All channels are subject to independent and identically distributed Rayleigh fading, i$.$e$.$, their elements are distributed independently as $\setC \setN (0,1)$; the propagation model further includes pathloss and shadowing, whose parameters are in accordance with the 3GPP TR 36.828 network layout and channel model \cite[Annex~A]{3GPP12}. The UL/DL performance is obtained via Monte Carlo simulations over $10^{4}$ realizations of the marked HCPP $\Phi$ and of the Rayleigh channels (each realization can be interpreted as a random instance of the network).

\begin{figure*}[t!]
\includegraphics[scale=0.98]{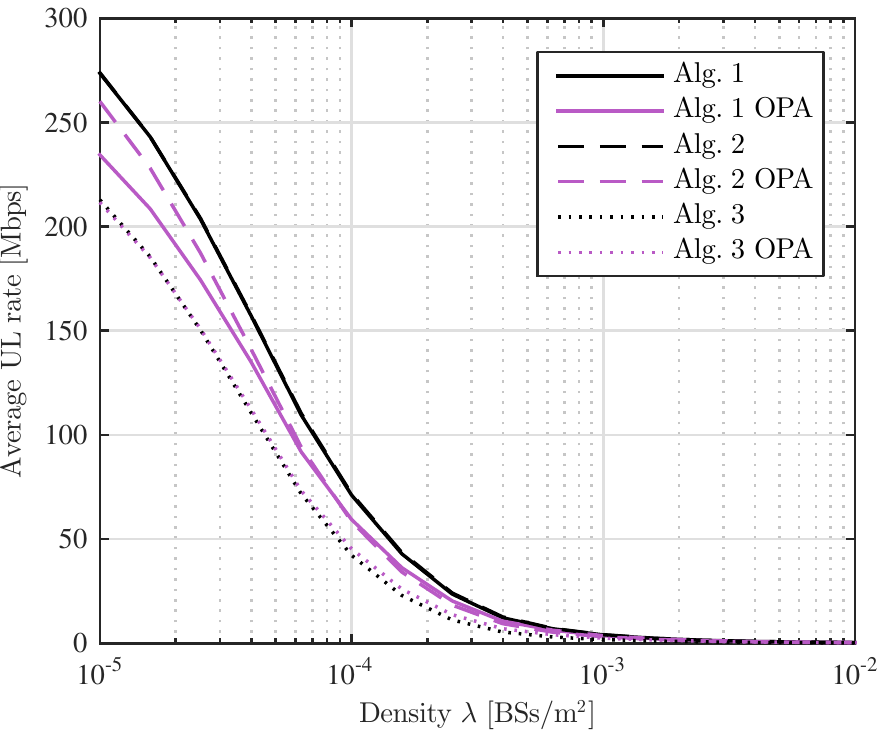} \hspace{3mm}
\includegraphics[scale=0.98]{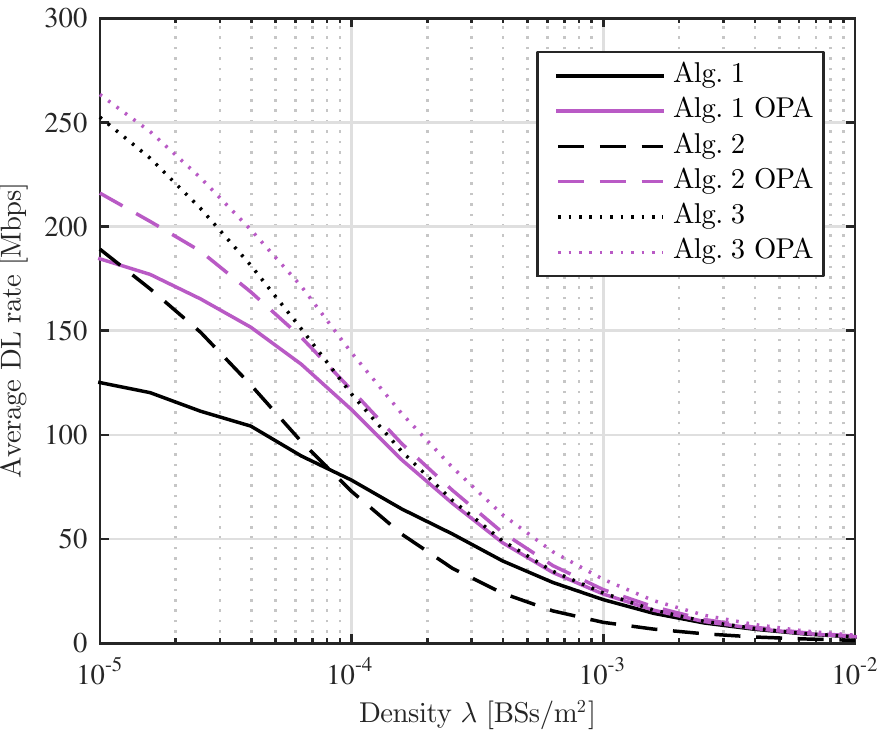}
\caption{Average UL and DL rate obtained with the different user-scheduling algorithms (with and without OPA), for $N_{T} \! = \! N_{R} \! = \! 1$ and different densities~$\lambda$.} \label{fig:dens_UL_DL} \vspace{-1mm}
\end{figure*}

\begin{figure}[t!]
\centering
\includegraphics[scale=0.98]{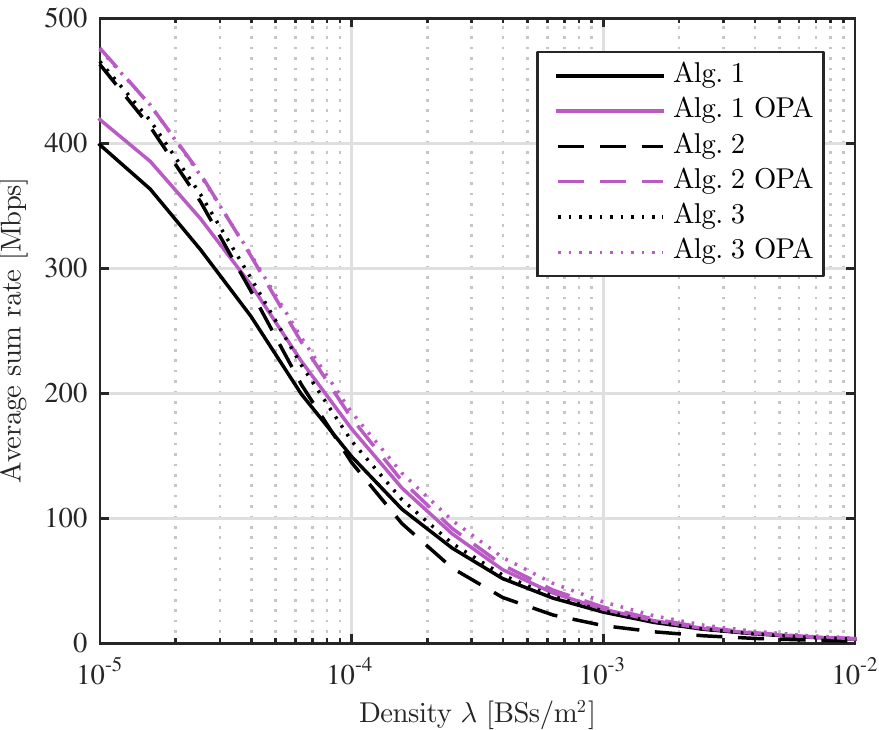}
\caption{Average sum rate obtained with the different user-scheduling algorithms (with and without OPA), for $N_{T}=N_{R}=1$ and different densities $\lambda$.} \label{fig:dens_sum} \vspace{-1mm}
\end{figure}

We set the following simulation parameters: $U_{0}=D_{0}=10$, $r_{0}=40$~m, $p_{\UL}=P_{\UL}=23$~dBm, $p_{\DL}=P_{\DL}=24$~dBm, $\Omega=110$~dB, and $B = 20$~MHz. As performance metrics to evaluate the proposed user-scheduling algorithms, with and without the OPA enhancement, we consider the UL and the DL rates $\mathsf{R}_{0}$ and $\mathsf{R}_{d_{0}}$ (see \eqref{eq:R_0} and \eqref{eq:R_d}, respectively). 

Considering the single-input single-output (SISO) case (i$.$e$.$, $N_{T}=N_{R}=1$) and $\lambda=2.5 \times 10^{-5}$~BSs/m$^{2}$, we plot the empirical CDFs of the UL and DL rates in Figure~\ref{fig:cdf_UL_DL} and of the sum rate in Figure~\ref{fig:cdf_sum}.\footnote{The density $\lambda$ used here corresponds to that of a macrocell network with inter-site distance of $500$~m and $6$~small-cell BSs per macro cell.} We observe that Alg.~1 performs better in the UL than in the DL, Alg.~2 improves the DL, whereas Alg.~3 sacrifices the UL for the DL. Furthermore, applying the OPA enhancement to Alg.~1 and Alg.~2 implies sacrificing the UL for the DL, whereas applying OPA to Alg.~3 improves the UL and the DL simultaneously. The best performance in terms of sum rate is obtained with Alg.~3; furthermore, as expected, the OPA enhancement improves the sum rate of all the algorithms. Figure~\ref{fig:hist_UL_DL} quantifies the contributions of the UL and DL SINRs (see \eqref{eq:SINR_0}--\eqref{eq:I_d_U}). As expected, selecting the UL and the DL MTs independently as in Alg.~1 introduces severe inter-MT interference, which impacts tremendously the DL performance and needs to be harnessed. On the other hand, the inter-MT interference is effectively handled using Alg.~2 and Alg.~3. In this regard, with respect to Alg.~1, we observe the following: \textit{i)} Alg.~2 maintains the same desired UL signal while slightly decreasing the desired DL signal and substantially reducing the inter-MT interference; \textit{ii)} Alg.~3 reduces the inter-MT interference as effectively as Alg.~2, although it benefits more the desired DL signal than the desired UL signal. A SIC capability $\Omega = 110$~dB at the receive chain of BS $0$ brings the SI almost at the same level as the noise; the interference produced by the interfering BSs and by their scheduled UL MTs is different in the UL and in the DL due to the adopted propagation model (see \cite[Annex~A]{3GPP12} for details). The average UL and DL rates and the average sum rate with variable density $\lambda \in [10^{-5},10^{-2}]$~BSs/m$^{2}$ are plotted in Figures~\ref{fig:dens_UL_DL} and \ref{fig:dens_sum}, respectively: in particular, Alg.~2 with OPA remarkably achieves a $19.1\%$ gain for $\lambda = 10^{-5}$~BSs/m$^{2}$.

\begin{figure*}[t!]
\includegraphics[scale=0.98]{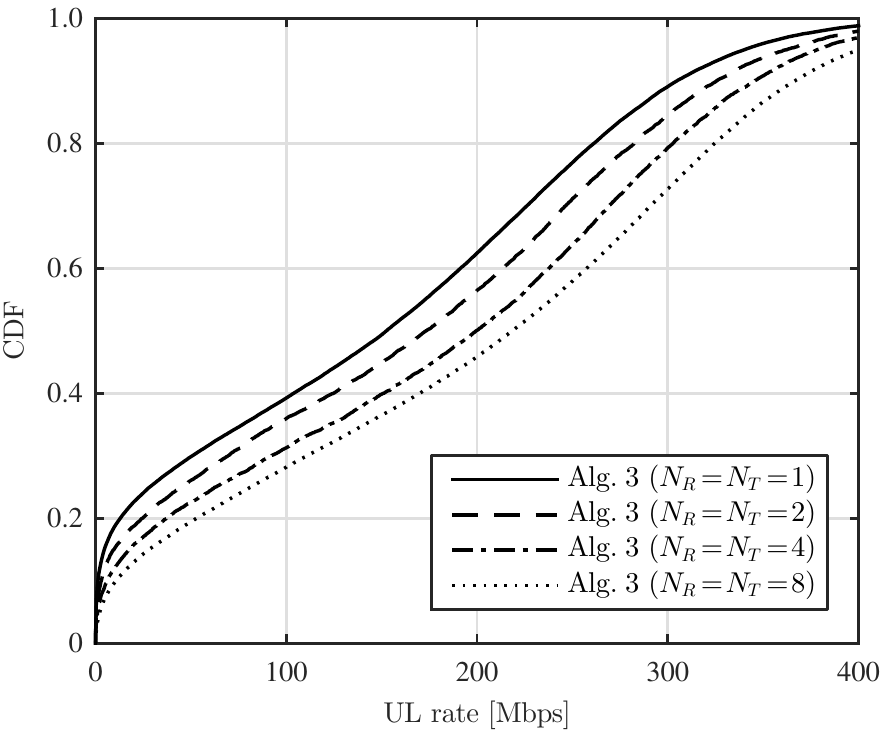} \hspace{3mm}
\includegraphics[scale=0.98]{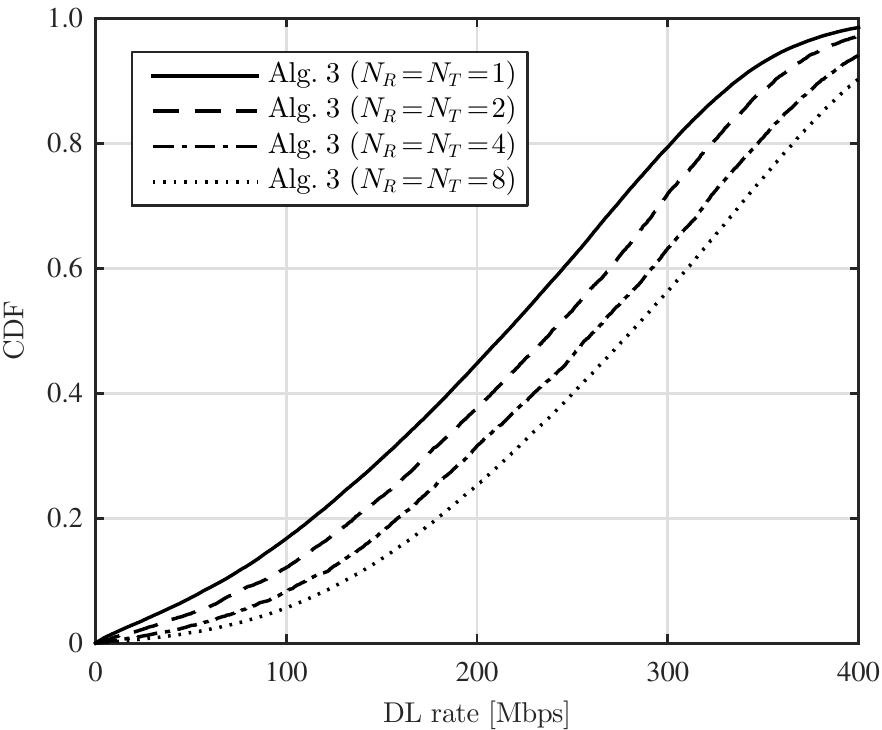}
\caption{Empirical CDFs of the UL and DL rate obtained with Alg.~3, for different number of antennas and $\lambda=2.5 \times 10^{-5}$~BSs/m$^{2}$.} \label{fig:cdf_LA} \vspace{-1mm}
\end{figure*}

\begin{figure}[t!]
\centering
\includegraphics[scale=0.98]{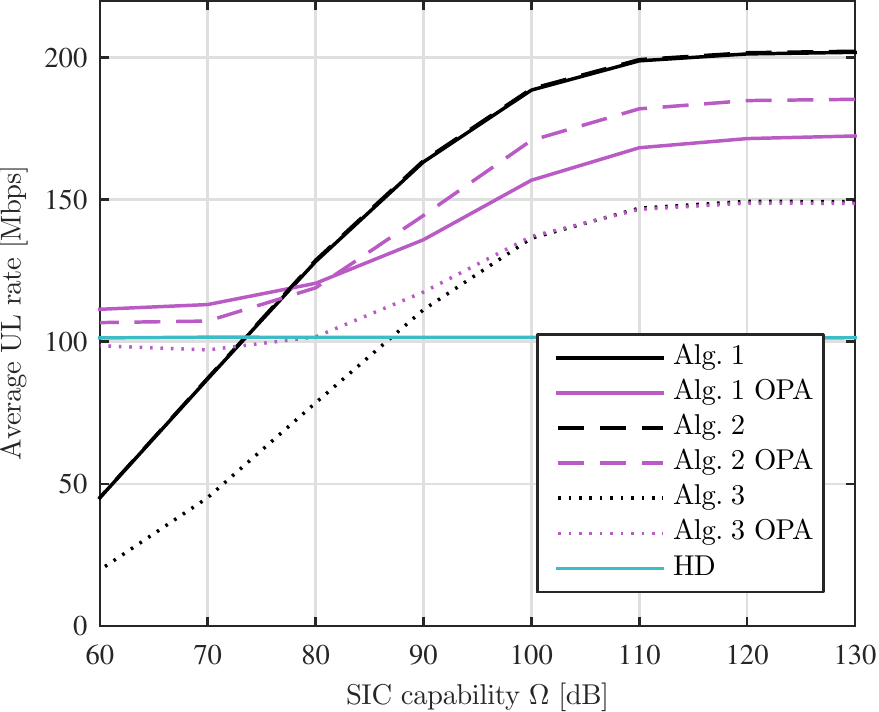}
\caption{Average UL rate obtained with the different user-scheduling algorithms (with and without OPA), for $N_{T}=N_{R}=1$, $\lambda=2.5 \times 10^{-5}$~BSs/m$^{2}$, and different values of the SIC capability $\Omega$ at the receive chain.} \label{fig:sic_UL} \vspace{-1mm}
\end{figure}

Now, setting again $\lambda=2.5 \times 10^{-5}$~BSs/m$^{2}$, Figure~\ref{fig:cdf_LA} considers Alg.~3, which is the user-scheduling algorithm that yields the best performance in terms of sum rate, in a multiple-input multiple-output (MIMO) setting and plots the empirical CDFs of the UL and DL rates: it is straightforward to observe how the employment of multiple antennas is beneficial for both UL and DL rate performance. Lastly, going back to the SISO case, we analyze the impact of the SIC capability $\Omega$ at the receive chain on the UL rate. In this respect, Figure~\ref{fig:sic_UL} plots the average UL rate for $\Omega \in [60,130]$~dB. We note that, below a certain threshold of $\Omega$, FD becomes infeasible and HD mode is optimal: in fact, all the user-scheduling algorithms with OPA converge to the HD regime.

\section{Conclusion} \label{sec:Concl}

In this paper, we evaluate the system-level performance of FD small cells in UDNs and study the impact of resource allocation schemes. More specifically, we consider: \textit{i)} three low-complexity user-scheduling algorithms, and \textit{ii)} power allocation between FD and UL/DL HD modes. Using spatial stochastic models for the network layout and 3GPP channel models, we show that simple cell-wide resource allocation schemes can yield significant throughput gains in different scenarios. Furthermore, we provide useful insights into the effect of multiple antennas, SIC capability, and BS density on the network performance.


\addcontentsline{toc}{chapter}{References}
\bibliographystyle{IEEEtran}
\bibliography{IEEEabrv,ref_Huawei}

\end{document}